\documentstyle[prl,aps,multicol,epsfig]{revtex}

\input epsf.tex

\newcommand{\be}{\begin{equation}}
\newcommand{\ee}{\end{equation}}
\newcommand{\ba}{\begin{eqnarray}}
\newcommand{\ea}{\end{eqnarray}}
\newcommand{\ban}{\begin{eqnarray*}}
\newcommand{\ean}{\end{eqnarray*}}

\newcommand{\braket}[2]{\mbox{$ \langle #1 | #2 \rangle $}}
\newcommand{\moy}[1]{\langle #1 \rangle}

\newcommand{\ket}[1]{\mbox{$ | #1 \rangle $}}

\newcommand{\demi}{\frac{1}{2}}
\newcommand{\compl}{\begin{picture}(8,8)\put(0,0){C}\put(3,0.3){\line(0,1){7}}\end{picture}}

\newcommand{\one}{\leavevmode\hbox{\small1\normalsize\kern-.33em1}}

\begin{document}

\title{Energy-time entanglement of quasi-particles in solid-state devices}
\author{Valerio Scarani$^1$, Nicolas Gisin$^1$, Sandu Popescu$^{2}$}
\address{$^1$Group of Applied Physics, University of Geneva, 20, rue de
l'Ecole-de-M\'edecine, CH-1211 Geneva 4, Switzerland\\
$^2$ H.H. Wills Physics Laboratory, University of Bristol, Tyndall
Avenue, Bristol BS8 1TL, UK}
\date{\today}
\maketitle

\begin{abstract}
We present a proposal for the experimental observation of {\em
energy-time entanglement} of quasi-particles in mesoscopic
physics. This type of entanglement arises whenever correlated
particles are produced at the same time and this time is uncertain
in the sense of quantum uncertainty, as has been largely used in
photonics. We discuss its feasibility for electron-hole pairs. In
particular, we argue that the recently fabricated 2DEG-2DHG
junctions, irradiated with a continuous laser, behave as
"entanglers" for energy-time entanglement.
\end{abstract}

%\pacs{PACS Nos. 03.65.Bz}

\begin{multicols}{2}

Entanglement lies at the heart of quantum mechanics, whose
astonishing features come mainly from it \cite{eprBohr}. Interest
on entanglement has grown, since it was recognized as a ressource
needed to perform tasks that would classically be impossible
\cite{QIPIntro98}. Correlative to the notion of entanglement is
the notion of {\em sub-systems}: algebraically,
$\compl^2\otimes\compl^2$ is equivalent to $\compl^4$, so, if one
cannot address separately the two sub-systems, one cannot
investigate entanglement. This is the reason why, although
entangled states arise in every sub-field of quantum physics (e.g.
the eigenstates of total momentum), it is usually an experimental
challenge to achieve control over entanglement. After the
spectacular results of photonics \cite{tittel}, entanglement has
recently demonstrated in other physical systems \cite{others}.
There is a growing list of proposals aimed at the observation of
entanglement in solid-state physics, using quantum dots, Josephson
junctions and other devices. In this Letter we focus on
quasi-particles in mesoscopic devices.

Coherent transport of quasi-particles in semi-conductors has been
widely demonstrated, so one can envisage to demonstrate
entanglement. A few years ago, Burkard and coworkers noticed that
electron-electron entanglement in spin could be detected by
measuring correlation in the current noise \cite{burkard}. In Ref.
\cite{impossible}, the scheme was completed with a proposal for an
"entangler", that is, for a source of spin-entangled electrons: a
Cooper pair from a superconducting material. Recently, these ideas
have been extended to entanglement in spatial degrees of freedom,
for two electrons generated as Cooper pairs \cite{buttiker} and
for electron-hole pairs in edge states \cite{beenakker}.

\begin{center}
\begin{figure}
\epsfxsize=7cm \epsfbox{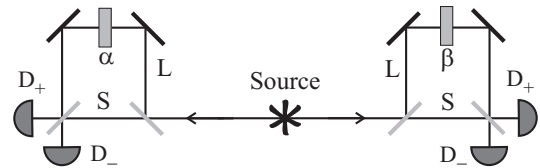} \caption{The Franson
interferometer, drawn for photonics. Grey segments are 50-50
couplers, $\alpha$ and $\beta$ are dephasing elements (small
delays). The difference between the two arms on each side, $\Delta
L = L-S$, must satisfy requirement (\ref{cond}).} \label{fig1}
\end{figure}
\end{center}

In the present Letter we take a different approach, that works in
principle for {\em any kind of particles produced in pairs}. The
basic idea is the following: two particles $a$ and $b$ are
produced at the same, but uncertain, time. This quantum
uncertainty is within the coherence time of the source. The latter
is typically a photon from a laser beam, called the pump photon,
whose well-determined energy is shared between the two produced
particles. Hence the energy and the time of creation of each
particle are uncertain, but the sum of the energies and the
difference of the times are well-determined. This form of
entanglement is known as {\em energy-time entanglement}. To
observe a signature of this entanglement for photon pairs, Franson
proposed in 1989 a very convenient interferometer
\cite{franson89}, sketched in Fig. \ref{fig1}. Each particle is
sent through unbalanced interferometers, with the same difference
between the long (L) and the short (S) arm:
$L_a-S_a=L_b-S_b=\Delta L$. If $\Delta L$ is larger than the
single-particle coherence lengths $\ell_c^{a,b}$, no
single-particle interferences will be observed. However, the
coherence length $\ell_c$ of the pair is usually much larger than
$\ell_c^{a,b}$. Let then \ba \ell_c\,>&\,\Delta L\,
&>\,\ell_c^{a,b}\,.\label{cond} \ea In this case, the alternatives
"both particles have taken the long arm" (LL) and "both particles
have taken the short arm" (SS) are indistinguishable and exhibit
interference fringes; while the two other alternatives, LS and SL,
are distinguishable because one particle clearly arrives before
the other one to its detector. Thus, in the runs in which both
particles are detected at the same time, an interference pattern
is observed that is due to the entanglement in energy-time.

{\em The photonic setup.} We start by briefly reviewing the
photonic setup, that has already been the object of successful
experiments e.g. \cite{brendel}. A non-linear crystal is pumped by
a cw laser with coherence time $\tau_c$. This coherence time is
defined as usual: the state of the laser light is a coherent beam
with a fluctuating phase $\phi(t)$, such that
$\phi(t+\tau)-\phi(t)$ is a stationary Gaussian process of mean
value $\moy{\phi}=0$ and of variance $\moy{\phi^2}=2\tau/\tau_c$.
A non-linear, purely quantum-mechanical process (parametric
down-conversion) takes place, that produces a field in two
initially empty modes $a$ and $b$, whose wave-vectors and
polarizations are determined by energy and momentum conservation.
If the intensity of the pump is weak enough, the field in $a$ and
$b$ consists essentially of a large vacuum component (that we
neglect) plus a two-photon component. In each mode $a$ or $b$, we
shall write $\ket{1,0}$ (resp., $\ket{0,1}$) for one photon
propagating along a horizontal (resp. vertical) direction in Fig.
\ref{fig1}. The state of the down-converted field can be written
as a superposition of two-photon fields produced at any time $t$:
\ba \ket{\Psi}& = & \sqrt{A}\,\int
dt\,e^{i\phi(t)}\,\ket{1_t,0}_a\ket{1_t,0}_b\,, \label{input}\ea
where $A$ is proportional to the power of the laser and the
efficiency of the down-conversion process. The states
$\ket{1_t,0}_{a,b}$ can be seen as an over-complete set; the
overlap $\braket{1_t,0}{1_{t'},0}_{a,b}$ decreases rapidly as a
function of $|t-t'|/\tau_c^{a,b}$, where $\tau_c^{a,b}$ are the
single-photon coherence time. As we discussed, in our experiment
this time is much shorter than the other times involved ($\Delta
t$, $\tau_c$).

The state (\ref{input}) can be seen as a continuous version of the
maximally entangled state of two $d$-dimensional quantum systems,
indexed by the parameter $t$. Franson's setup is a way of
partially detecting this entanglement, by projection onto a
two-dimensional subspace and post-selection. The evolution of mode
$a$ in the unbalanced interferometer --- the beam-splitters are
50-50 couplers --- is \ba \ket{1_t,0}_a &\longrightarrow &
\ket{1_t,0}_a\,+\, i
\ket{0,1_t}_a\,+\nonumber\\
&& + ie^{i\alpha} \ket{0,1_{t+\Delta t}}_a -\, e^{i\alpha}
\ket{1_{t+\Delta t},0}_a
 \label{evolph}\ea where we have omitted a global factor $\demi$ and have re-defined the origin of
time to take into account the propagation from the source. The
evolution of mode $b$ is identical, with a phase $\beta$ instead
of $\alpha$.

The two-photon state at the detection stage is obtained by
replacing the evolved state into (\ref{input}): it is a sum of
sixteen basic kets. We focus on a pair of detectors, say the two
detectors labelled $+$ in Fig. \ref{fig1}. This means that we
project onto the four kets of the form $\ket{1,0}_a\ket{1,0}_b$,
that we write for conciseness $\ket{1;1}$: \ba
\ket{\Psi_{++}}&\simeq &\int
dt\,e^{i\phi(t)}\,\Big[\ket{1_t;1_t}\,+\,
e^{i(\alpha+\beta)}\ket{1_{t+\Delta
t};1_{t+\Delta t}} +\nonumber\\
&&+\,e^{i\beta}\ket{1_t;1_{t+\Delta t}}\,+\,
e^{i\alpha}\ket{1_{t+\Delta t};1_t}\Big]\,. \ea The two first
terms correspond to the cases where the two photons arrive at the
same time in the detectors (paths SS and LL in Fig. \ref{fig1});
because of the invariance through translation in time, these two
cases are indistinguishable, and interfere. We can re-write these
two first terms as \ba \ket{\tilde{\Psi}_{++}}&\simeq &\int
dt\,\big[e^{i\phi(t)}\,+\, e^{i(\alpha+\beta+\phi(t-\Delta
t))}\big]\ket{1_t;1_t}\,. \label{interf}\ea The third term is the
case where photon $b$ is delayed by $\Delta t$ with respect to
photon $a$ (path SL), and the fourth term is the opposite case
(path LS). These last two cases are {\em in principle}
distinguishable from the others, so they don't contribute to any
interference \cite{note1}. It is a different matter of course to
distinguish them {\em in practice}. So, a priori we have to
consider two possible outcomes of the experiment: if one can
select only the interfering cases, the detection rate is
\cite{note5} \ba \tilde{R}(++)&=& ||\tilde{\Psi}_{++}||^2\,
\propto \, 1 + e^{-\frac{\Delta t}{\tau_c}}\cos(\alpha+\beta)\,;
\label{vis1}\ea recalling that the visibility $V$ is defined by
$R\propto 1+V\cos\theta$ for a sinusoidal fringe, we find $V=
e^{-\frac{\Delta t}{\tau_c}}\approx 1$. If one cannot select only
the interfering cases, the visibility of the observed interference
fringes will be reduced down to $V\approx \demi$, since one has
\ba R(++)&=&||{\Psi}_{++}||^2\, \propto \, 2 + \,e^{-\frac{\Delta
t}{\tau_c}}\cos(\alpha+\beta)\,. \label{visdemi}\ea In optics, for
typical coherence times and jitters of the detectors, one {\em
can} select only those cases where the photons arrive at the same
time; that's how $V\approx 1$ has been reached and the Bell
inequality could be violated \cite{brendel}.

{\em The proposal: overview.} We can now turn to the main goal of
this paper: a proposal for the production and detection of
energy-time entangled quasi-particles in mesoscopic physics
\cite{note2}. Specifically, we consider {\em electron-hole pairs}
produced in semiconductor junctions illuminated by a laser. A low
intensity cw laser with coherence time $\tau_c$ illuminates a
junction, producing electron-hole pairs. When the electron and the
hole do not recombine, they will be accelerated out of the
junction in opposite directions. Once produced, each
quasi-particle travels in a semi-conductor structure, tailored for
single-mode coherent transport of the electron \cite{cohel} or the
hole \cite{cohol}: typically, a two-dimensional electron or hole
gaz, noted respectively 2DEG and 2DHG. The unbalanced Mach-Zehnder
interferometer is engineered in the semi-conductor, in the form of
an asymmetric loop, where the phase between the two arms can be
varied using the Aharonov-Bohm effect \cite{machzehnd}. The two
paths are then recombined and split again, each ending in a
detector. The rest of the paper is devoted to a detailed analysis
of the three parts of the setup: entangler (preparation),
interferometer (evolution), and detectors (measurement).

\begin{center}
\begin{figure}
\epsfxsize=9cm \epsfbox{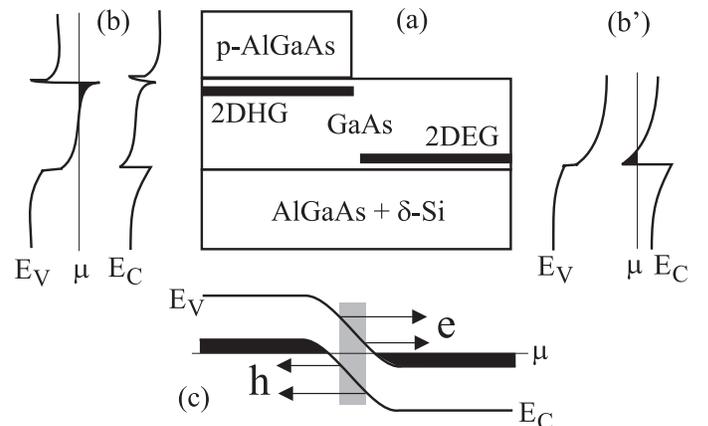} \caption{Qualitative
description of the 2DEG-2DHG junction. (a) The design. (b,b') The
gap along the vertical axis. (c) Schematic description of the
junction and of its behavior when irradiated with a laser spot
(grey area).} \label{fig3}
\end{figure}
\end{center}

{\em The entangler.} A standard bulk p-n junction is most probably
not an entangler: because of the massive doping, the motion of the
quasi-particles in the device is in principle diffusive rather
then ballistic, implying that quantum phase coherence is lost even
before reaching the interferometers. Even if this description was
too pessimistic, one should bother about the interfaces between
the bulk and the 2D-gases. Fortunately, there is an elegant way of
by-passing both obstacles: by creating a {\em junction between a
2DEG and a 2DHG}, the source is just the interface, the
interferometer can be engineered in the same materials and the
whole motion can be ballistic. The first 2DEG-2DHG junction has
been fabricated very recently in AlGaAs/GaAs heterostructures
\cite{kaestner} according to the scheme that we reproduce in Fig.
\ref{fig3}. The full understanding of the physics of such
junctions and the optimization of the parameters will need further
work \cite{note0}. But there is no fundamental objection to
considering a 2DEG-2DHG irradiated by a laser behave as an
entangler to generate electron-hole pairs entangled in
energy-time. Even more, this goal may be a strong motivation to
boost technical improvements. Finally note that p-n junctions have
also been fabricated in another material exhibiting ballistic
transport, namely carbon nanotubes \cite{nanotubes}. These can
also be candidates as entanglers.

{\em The interferometer.} From now on, we shall use for numerical
estimates typical values for electrons, extracted from Ref.
\cite{datta}. One must not forget that holes often have smaller
mobilities in these structures, so the figures may not apply to
one half of the interferometer --- but the principles of the
analysis do apply. In the forthcoming discussion, we shift when
convenient from "lengths" $\ell$ to "times" $\tau$, the link being
provided by $\ell = v_F\tau$ where $v_F$ is the Fermi velocity in
a 2DEG or 2DHG, typically $v_F \sim 3\times 10^{7}\mbox{ cm/s}$.
In the optical experiment, we introduced the requirement
(\ref{cond}) on $\Delta L$ for the Franson interferometer to show
two-particle interferences. In the present proposal, it is trivial
to have the coherence length of the pair $\ell_c$ exceed all the
other meaningful lengths, since $\ell_c$ is determined by the
coherence time of the pump laser, and cw lasers easily have a
coherence time of tens of nanoseconds. However, here we must meet
an additional constraint due to the {\em role of the environment}.
While photons essentially do not couple to the environment,
electrons and holes propagating in semi-conductors interact
strongly, especially with the other quasi-particles. Because of
this coupling, some which-path information is transferred out of
the system under study, whose coherence is thus decreased. So the
size of each interferometer must not be too big: for both
electrons and holes, $L+S$ should be much smaller than
$L_{\varphi}$, a phase-relaxation length that characterizes the
coupling with the environment. Since (at least in principle) the S
path can be made arbitrarily short we have $L+S\approx \Delta L$,
and we can summarize our present requirements as
$L_{\varphi}^{e,h}>\,\Delta L^{e,h} \,> \ell_c^{e,h}$. Confident
in the precision of semiconductor manufacturing techniques, we
admit that this requirement holds if and only if
$L_{\varphi}^{e,h}>>\ell_c^{e,h}$, that is, if \ba
\tau_{\varphi}^{e,h}&>>& \tau_c^{e,h}\,. \label{req1}\ea We focus
on the electron, the analog holds for the hole.

We want to show that (\ref{req1}) can be fulfilled if the electron
is injected into the 2DEG close enough to the Fermi level. Refer
to Fig. \ref{fig3} (c). The electron is injected into the 2DEG
with an energy $E=\epsilon_F+\Delta\epsilon$, where $\epsilon_F$
is the Fermi energy of the 2DEG, typically some 0.1 eV
\cite{datta}. Since the main relaxation mechanism will be e-e
inelastic scattering, the phase-relaxation time is
$\tau_\varphi\sim \hbar\epsilon_{F} / \Delta\epsilon^{2}$
\cite{theory}. For instance, if the electron is injected into the
2DEG with $\Delta\epsilon=\frac{1}{100}\epsilon_F\sim 10^{-3}$ eV,
then we obtain $\tau_\varphi^{e}\simeq 100$ ps, in good agreement
with the observed values of $L_{\varphi}^{e,h}$, that are
typically some 10 $\mu$m (see \cite{cohol} and refs therein). The
single-particle coherence time can be estimated by $\tau_c^{e}\sim
\frac{\hbar}{\Delta E}$, where $\Delta E$ is the uncertainty in
the electron kinetic energy. In Fig. \ref{fig3} (c), one sees
clearly that this uncertainty is determined by the relation
between the steepness $\phi'=\frac{d\phi}{dx}(x=0)$ of the
built-in potential and the width $w$ of the laser spot. In
particular, $\Delta E\lesssim (\Delta\epsilon)_{max} \approx
e\phi'\,w$, this being the largest value of $\Delta\epsilon$. From
the expressions of $\tau_\varphi^{e}$ and $\tau_c^{e}$, one
derives immediately that condition (\ref{req1}) holds if and only
if $(\Delta\epsilon)_{max}<<\epsilon_F$, and this can in principle
be achieved by decreasing $\phi'$ (junction engineering) or $w$
(laser focusing).

Condition (\ref{req1}) being possible, the calculation of the
two-particle interference pattern goes along the same lines as for
the optical implementation \cite{note3}, adding the presence of
different environments. Following the approach of Stern et al.
\cite{stern}, we write $\ket{\epsilon_L^{e}}$,
$\ket{\epsilon_S^{e}}$, $\ket{\epsilon_L^{h}}$ and
$\ket{\epsilon_S^{h}}$ the four environments associated to the
paths, where $L,S$ stand for Long and Short, and $e,h$ stand for
electron and hole. The evolution of the electron state, replacing
(\ref{evolph}), is then \ba \ket{1_t,0}_e &\longrightarrow &
\big(\ket{1_t,0}_e\,+\, i
\ket{0,1_t}_e\big)\,\ket{\epsilon_S^e}\,+\nonumber\\
&& + \big(ie^{i\alpha} \ket{0,1_{t+\Delta t}}_e -\, e^{i\alpha}
\ket{1_{t+\Delta t},0}_e\big)\,\ket{\epsilon_L^e} \label{evole}\ea
and a similar evolution for the hole. In the interfering terms
$\ket{\tilde{\Psi}_{++}}$, the term into brackets in eq.
(\ref{interf}) is replaced by \ba
e^{i\varphi(t)}\ket{\epsilon_S^e} \ket{\epsilon_S^h}\,+\,
e^{i(\alpha+\beta+\varphi(t-\Delta t))}\ket{\epsilon_L^e}
\ket{\epsilon_L^h}\,\ea and finally the visibility is reduced with
respect to the analogous optical visibility as \ba V_{e-h}&=&
V_{opt}\,\big|\braket{\epsilon_S^e}{\epsilon_L^e}\big|\,
\big|\braket{\epsilon_S^h}{\epsilon_L^h}\big|\,. \label{visi}\ea
The phase-relaxation length that we introduced above is related to
the expressions in (\ref{visi}), in the simplest case
\cite{stern}, through $\big| \braket{\epsilon_S^x}{\epsilon_L^x}
\big|= e^{-(L_x+S_x)/L_{\varphi}^{x}}$ for both $x=e,h$.

{\em The detectors.} A convenient detection scheme should use the
observables of mesoscopic physics, that are current-correlation
measurements \cite{buttiprl}. We can consider that the detector
for the electron, resp. the hole, is a metal reservoir biased with
a voltage $-V$, resp. $+V$. If the temperature is small enough
($kT<<|eV|$), no electron (hole) can be injected from the
reservoirs into the semiconductor. The situation becomes then
analog to those studied in Refs.
\cite{burkard,impossible,buttiker,beenakker}: entanglement can be
detected by measuring the zero-frequency current cross-correlator.
Since this detection scheme is not time-resolved, it leads to
(\ref{visdemi}) with the correction (\ref{visi}). We want to
conclude by addressing the question of time-resolved detection,
that is, the hope of observing the analog of (\ref{vis1}). Typical
for energy-time entanglement is the presence of two meaningful
timescales at the detection. The first, standard one, comes from
the fact that want at most a single electron-hole pair to be
produced per coherence time of the pump $\tau_c$. This timescale
determines the current: taking $\tau_c\simeq 1$ ns, large enough
to ensure $\tau_c>>\tau_{\varphi}^{e,h}\simeq 100$ ps, one expects
a current at detection of some 100 pA. The second timescale
determines whether one can discriminate the interfering cases from
the non-interfering ones. Explicitly, a time resolution for the
single-electron (-hole) measurement of $\tau_{meas}<\Delta
L/v_F\sim 10$ ps is needed to enter the regime $V>\demi$
(\ref{vis1}). Both time-resolution and sensitivity are in
principle achievable using single-electron transistors as
detectors \cite{devoret}. But very fast measurements introduce
unwanted excitations that may hide the signal; moreover,
$\tau_{meas}>\tau_c^{e,h}$ must hold in order to detect the
quasi-particles. Anyway, $V\approx \demi$ would already be a fair
demonstration of entanglement, because the origin of this reduced
visibility in a Franson setup is well understood.

In conclusion, we have argued that energy-time entanglement of
quasi-particles can be observed. The task is challenging, but the
goal seems within reach with present-day technology. 2DEG-2DHG
junctions are promising candidates as entanglers for energy-time
entangled electron-hole pairs.

We thank M. B\"uttiker, P. Samuelsson, E.V. Sukhorukov and B.
Kaestner for useful comments. V.S. acknowledges financial support
from the Swiss NCCR "Quantum Photonics", as well as fruitful
discussions with members of the network.

\end{multicols}


\begin{thebibliography}{99}
\bibitem{note1} In our formalism, this qualitative reasoning amounts to set
$\braket{1_t,0}{1_{t+\Delta t},0}_{a,b}=0$, that is, to neglect
the single-particle coherence times $\tau_c^{a,b}$ before $\Delta
t$.
\bibitem{note5} For the calculation of both (\ref{vis1})
and (\ref{visdemi}), we use $E
\big[e^{i(\phi(t)-\phi(t-\xi))}\big]=e^{-|\xi|/\tau_c}$, where
$E[\cdot]$ is the expectation value of the random variable.
\bibitem{note2} In particular, since there is only one
particle per mode in a Franson interferometer, the bosonic or
fermionic nature of the particle does not introduce any
difference.
\bibitem{note0} For instance, in the device of Ref. \cite{kaestner}, the 2DEG and 2DHG are
separated vertically by 90nm; this means that the electron and the
hole, before being coupled in the 2D-gases, must travel some tens
on nanometers in the intrinsic GaAs, where the phase coherence
time is shorter. Also, it would be useful to decrease the
steepness of the built-in potential (see main text, below).
\bibitem{note3} We admit that the two splittings of each
semi-conductor structure are 50-50 beam-splitters.
\end{thebibliography}
\end{document}